\documentclass[11pt]{article}
\usepackage{epsfig}
\usepackage{amssymb,amsmath}
\usepackage{epigraph}

\usepackage{graphicx}				
\usepackage{amssymb}

\newcommand{\beq}{\begin{equation}}
\newcommand{\eeq}{\end{equation}}
\newcommand{\bea}{\begin{eqnarray}}
\newcommand{\eea}{\end{eqnarray}}

\begin{document}



\begin{center}

{\LARGE
The Inverted Parabola World of Classical Quantitative Finance:
\vskip0.5cm
Non-Equilibrium and Non-Perturbative Finance Perspective  
}

\vskip1.0cm
{\Large Igor Halperin} \\
\vskip0.5cm
NYU Tandon School of Engineering \\
\vskip0.5cm
{\small e-mail: $ighalp@gmail.com $}
\vskip0.5cm
\today \\

\vskip1.0cm
{\Large Abstract:\\}
\end{center}
\parbox[t]{\textwidth}{
Classical quantitative finance models such as the Geometric Brownian Motion or its later extensions such as local or stochastic volatility models do not make sense when seen from a physics-based perspective, as they are all equivalent to a negative mass oscillator with a noise.
This paper presents an alternative formulation based on insights from physics. 
}

 \newcounter{helpfootnote}
\setcounter{helpfootnote}{\thefootnote} 
\renewcommand{\thefootnote}{\fnsymbol{footnote}}
\setcounter{footnote}{0}
\footnotetext{
I would like to thank Peter Carr for critical remarks. All possible errors are my own.
}     

 \renewcommand{\thefootnote}{\arabic{footnote}}
\setcounter{footnote}{\thehelpfootnote} 

\newpage
 
\section{Introduction}

The novel "The Inverted World" by Christopher Priest paints a fascinating image of a world where a city called the City of Earth
 slowly travels on railway tracks across an alien planet. The city's engineers keep laying a fresh track for the city, and pick up the old track as it moves. 
The city must move on to stay within 10 miles of the Optimum - which is a location where the gravitational field is not distorted, and matches the gravitational field of the planet Earth. 

But the world of the novel is not Earth. In this world, the ground is in a constant move from the north to the south, as a result of some sort of a global gravitational catastrophe that happened at some point in the past. Even as the Optimum stays in the same position, the City will drift away from the Optimum if it does not move all the time. If the City of Earth finds itself too far from the Optimum, gravitational distortions become too strong as a result of moving grounds and their drift to the south. This is what makes the City crawls to the North all the time. If it ever stops, it will eventually be pulled to the South, and destroyed at the end by gravitational distortion forces, along with all its citizens. So it has to move forward  through a devastated land full of hostile tribes. The only alternative to a constant move in such an inverted world, where the grounds are moving and the Sun looks like a rotating parabola, is death.

The resolution of the many puzzles and gaps of history of the Inverted World comes only towards the end of the book. The City of Earth was crawling the planet of Earth, never leaving it.
Moving grounds, a parabolic Sun, and other related puzzles of the Inverted World was caused by side neurological effects of a UV radiation
that was produced by the city's power generator. The generator was  based on an alternative energy method that was developed by a founder of the city.     
 
This novel, which I first read many years ago, came repeatedly to my mind when I worked on a model of asset price dynamic in an open and non-equilibrium market called the Quantum Equilibrium-Disequilibrium (QED) model \cite{QED}. The QED model generalizes the Geometric Brownian Motion (GBM) model by introducing two additional parameters, along with a non-linear extension of a diffusion equation driving the dynamics of the model, see also \cite{MLF}. Surprisingly, the QED model suggests that the notions of a market growth  and market stability in this model and in the GBM model are essentially {\it opposite}. 

What was {\it 'equilibrium'} dynamics in the GBM model becomes {\it non-equilibrium} dynamics from the perspective of the QED model. What was the commonly excepted average exponential {\it growth} of asset prices becomes a {\it fall} from a point of instability  towards a point of local stability. This sounds much like an Inverted World vs a Normal World, the only question is which one is the Inverted World? 

This paper offers a non-technical introduction to the QED model, along with a reasoning why it corresponds to a 'Normal World', while the GBM model, along with its multiple direct descendants such as local or stochastic volatility models, describes an 'Inverted World'. 
As I will try to argue, assumptions of closed-system dynamics, (quasi-)stationarity and linearity made in classical financial models do not adequately capture realities of real world financial markets, and in a sense can be viewed as `wrong limits' of a (yet unknown) `right' theory. 
Then I show how these deficiencies are addressed in the QED model that treats markets as open and non-linear systems, and does \emph{not} rely 
on a linearization of dynamics within a perturbation theory to treat non-linearities. Instead, the QED model presents a \emph{non-perturbative} approach to handle 
non-linearities.
This position paper discusses how insights from modern non-equilibrium and non-perturbative physics can be fruitfully used for financial modeling with non-linear and non-equilibrium models such as the QED to better capture the true market dynamics.

\section{GBM, Langevin equation, and Inverted Parabola} 
\label{GBM_Parabola}

\subsection{GBM and the Langevin equation}
\label{sect:GBM_and_Langevin}

Since the groundbreaking work of Samuelson in 1965 \cite{Samuelson}, Geometric Brownian Motion (GBM) model, also known as the   
the log-normal asset return model,
 \beq
 \label{GBM}
 d X_t  =  \mu X_t dt    +  \sigma X_t  d W_t
\eeq
remains the main work-horse of financial engineering. In Eq.(\ref{GBM}), $ X_t $ is an asset price at time $ t $, $ \mu $ is the stock drift, $ \sigma $ is the stock volatility, and 
$ W_t $ is a standard Brownian motion. For what follows, we can view  the GBM model as a special linear case of a more general model called It{\'o}'s diffusion
 \beq
 \label{Itos_diff}
 d X_t  =  \mu (X_t) dt    +  \sigma (X_t)  d W_t
\eeq
with a {\it linear} drift function $ \mu(X_t) =  \mu X_t  $ and a {\it multiplicative} (i.e. proportional to $ X_t $) diffusion function $ \sigma(X_t) = \sigma X_t $.  
Samuelson proposed the GBM model (\ref{GBM}) as an improvement over an Arithmetic Brownian Motion (ABM) model suggested by Bachelier in 1900 \cite{Bachelier}.  His objective was to modify the ABM model to ensure non-negativity of stock prices. Note that the ABM model itself can be viewed as a model with a {\it constant} drift and volatility terms. 

%
%

A few years after Bachelier published his 1900 thesis that gave birth to the ABM model, Paul Langevin 
proposed in 1908 an equation that later became known as the Langevin equation. 
Langevin's work focused on a simplified analysis of overdamped Brownian particles within the Einstein-Smoluchovski theory of classical diffusion in the presence of an external potential field $ U(X) $.  Such a field can represent an impact of heavy molecules, a external gravitational or electromagnetic field, etc. \cite{Langevin}. 
The Langevin equation can be written in similar terms to It{\'o}'s diffusion (\ref{Itos_diff}), 
except that in the Langevin dynamics, a drift term is  given by the negative gradient of the 
potential $ U(X) $. The (overdamped) Langevin equation with a multiplicative noise reads
\beq
 \label{Langevin}
 d X_t  =  - \frac{\partial U(X_t)}{\partial X_t}dt    +  \sigma X_t  d W_t  
\eeq 
Therefore, the Langevin equation that is rooted in physics provides an interpretation to the drift term  
$ \mu(X_t) $ in the mathematical construction of It{\'o}'s diffusion (\ref{Itos_diff}):
any drift function $ \mu(X_t) $ can be viewed as a negative gradient of a potential $ U(X_t) $ in the equivalent Langevin dynamics (\ref{Langevin}). 
In particular, when the potential $ U(X_t) $ has a minimum, the Langevin equation describes a stochastic relaxation towards this minimum, where the gradient of the potential vanishes.



\begin{figure}[ht]
\begin{center}
\includegraphics[
width=80mm, 
height=50mm]{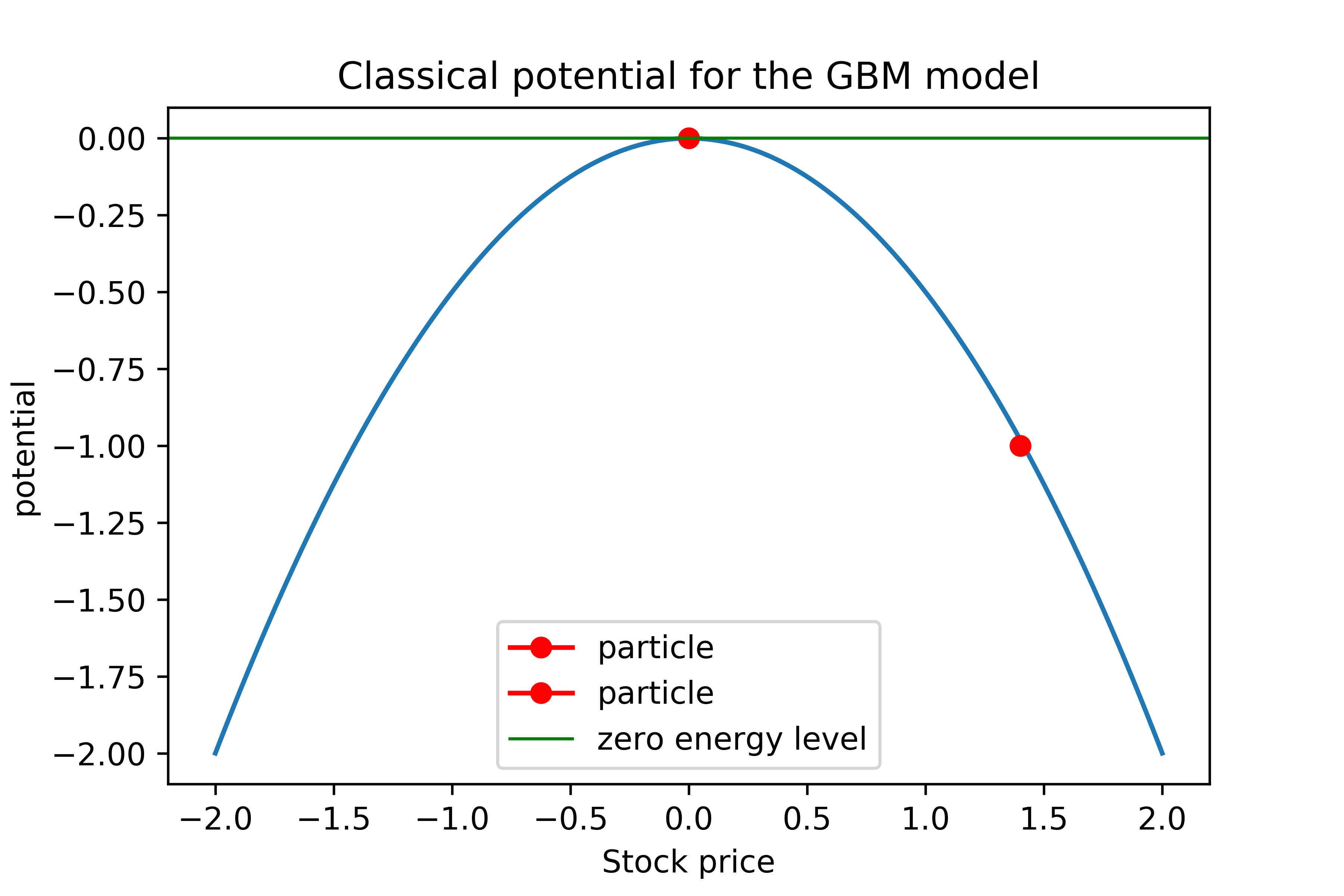}
\caption{The classical potential $ U(x) $ corresponding to the GBM model with $ \mu > 0 $. Red dots correspond to a ``particle" representing the firm, coordinate 
$ X_t $ being the firm's stock price.
This is the potential of a harmonic oscillator with a \emph{negative} mass. Such a system is globally unstable, and the default state $ X_t = 0 $ is unreachable as the force of the negative gradient of the potential pushes the particle away from the default boundary $ X_t = 0 $ for any value $ X_t > 0 $, producing an ever-accelerating
and unbounded fall in this potential.}
\label{fig:GBM_classic_potential}
\end{center}
\end{figure}

While this observation applies to any drift function $ \mu(X_t) $ in Eq.(\ref{Itos_diff}), it is of particular interest to explore its consequences for the  GBM model (\ref{GBM}).
 Comparing Eqs.(\ref{Langevin}) and (\ref{GBM}), we observe that the GBM model corresponds to a special case of a more general Langevin equation for the following choice of the force potential $ U(X) $:
 \beq
 \label{inverted_pot_GBM}
 U_{GBM}(X) = - \frac{\mu}{2}  X^2
 \eeq
 When the drift $ \mu $ is positive, this is the potential of an {\it inverted} harmonic oscillator with 'mass' $ \mu $!  Such a potential has a maximum at $ X = 0 $ and no minimum, see Fig.~\ref{fig:GBM_classic_potential}. 
From elementary physics, a classical motion in such potential describes an {\it unstable} system.\footnote{A similar analysis can be performed by changing to log-prices  $ y_t := \log X_t $. When expressed in terms of log-prices $ y_t $, the noise term becomes additive, while the $ y$-space potential is $ V(y) = - \left( \mu - \frac{\sigma^2}{2} \right) y $. When $ \mu > \frac{\sigma^2}{2} $, this potential corresponding to a uniform force pushing the particle away from the negative infinity 
$ y = - \infty $ corresponding to a financial distress or a bankruptcy. On the other hand, taking  $ \mu < \frac{\sigma^2}{2} $ would produce a linear attraction to $ y = - \infty $ and defaults (bankruptcies) happening too fast.} 

\subsection{The Inverted Parabola World of the GBM model and its descendants}
\label{sect_Inverted_parabola_world}



The inverted parabola potential (\ref{inverted_pot_GBM}) thus describes the most classical example of an unstable system in physics - an inverted (negative mass) harmonic oscillator. Of course, the fact that the GBM model is non-stationary for $ \mu \neq 0 $ is evident and well known in the literature. 
On the other hand, 'classical' financial inter-temporal models
(e.g.  the Black-Scholes model \cite{BS, Merton} or inter-temporal versions of the CAMP model \cite{CAPM}) often work under assumptions of a general equilibrium or competitive market equilibrium.
In these approaches, one assumes a dynamic market equilibrium between rational financial agents having instantaneous access to either symmetric or asymmetric information. Here the concept of a market equilibrium refers not to a price process, but rather to an equilibrium of supply and demand given a price level (recall the Equilibrium in "Inverted World").
Thus, the stock price is considered a 'reference frame' to describe the supply-demand balance equations. As the stock price dynamics is non-stationary, the same holds for the full system given by both financial agents (traders) \emph{and} price dynamics. This means that the concept of a market equilibrium under a non-stationary price process can be at best applied only approximately for short times, as a \emph{local approximation}.

To illustrate this point, imagine you step into an elevator on a top floor of a skyscraper. All of a sudden, the elevator cable breaks, and now the elevator is in a free fall, with you trapped inside.\footnote{Of course, this is largely a hypothetical scenario, see e.g. https://science.howstuffworks.com/science-vs-myth/everyday-myths/question730.htm regarding practical safety measures to prevent it from happening.} According to elementary physics, as long as the elevator continues to freely fall, you will be levitating inside of the elevator, being in a state of a local `equilibrium'. This is because being inside of a freely falling elevator is equivalent to residing in a non-inertial reference frame, where a fictitious 'anti-gravitational' force exactly cancels the gravitational force. This simple example illustrates the point that from the point of view of an external observer who observes \emph{both} the elevator and you inside of the elevator, the dynamics of the full system most definitely \emph{cannot} be described as equilibrium dynamics proceeding indefinitely in time. 
    
Far from being a purely theoretical observation, the inverted parabola potential (\ref{inverted_pot_GBM}) implies a completely absurd behavior of stock prices in the 
GBM model, as well as its all direct descendants such as local or stochastic volatility models, that becomes especially transparent in a small noise limit.\footnote{Analysis of a small noise limit is useful in order to not get 'fooled by randomness'. Note that this is \emph{not} the same as setting volatility $ \sigma $ to zero \emph{exactly}. In such a strict limit, a stock becomes a riskless asset that should earn a risk-free rate $ r $ according to a no-arbitrage argument, and thus should be the same as cash in a bank account.
As the exponential growth $ e^{r t} $ follows as a solution of a compound interest equation, financial mathematicians typically have no issue with taking the exponential law $ e^{r t} $ on its face value, and formally applying it for arbitrary times $ t \rightarrow \infty $ arguing that 'we believe this will continue in the next 100 years or so, given the past experience'. I believe that an appeal to the money bank account law  $ e^{r t} $ as a `justification' for an unbounded average exponential growth for stocks would be erroneous both financially and mathematically. It is wrong financially because stocks are \emph{not} cash, they can default, while a bank account is protected by state regulations. It is also wrong mathematically because the limit $ \sigma = 0 $ is a singular limit: corporate defaults become mathematically impossible in this limit, while their probability may remain small but non-zero for arbitrarily small but non-zero values $ \sigma > 0 $.  
} While the fact the the zero-price level $ X = 0 $ is not attainable in the GBM model is well known, the nature of this mechanism is rarely discussed. 
A critical observation is that not only the GBM model is incompatible with corporate defaults due to its inability to reach the zero level $ X = 0 $, but rather that non-negative prices are obtained in the GBM model at the cost of introducing a completely fictitious and \emph{absurd} force, due to the negative gradient of the GBM potential, that somehow \emph{saves} a firm from default once it gets close to the zero level. More than that, this force becomes unboundedly \emph{stronger} as the price increases! The presence of such an absurd ever-growing force for all positive prices appears to be too steep a price to pay for non-negativity of stock prices, which was the original motivation for the GBM model. 

Interestingly, the last observation that the repelling force actually \emph{increasses} rather than \emph{decreases} 
as the price moves away from the default boundary $ X_t = 0 $ 
also implies the model behavior should also become progressively less trustworthy for large values of $ X_t $, that could be expected in a  long run for the GBM dynamics (\ref{GBM}). The GBM model predicts that on average, the price of a given stock should grow exponentially in time, but 
empirically, a very few stocks have observable prices for a long period of, say, 100 years.\footnote{I would like to thank Peter Carr for pointing out one such stock: Sotheby's (BID).}
Most of stocks live much shorter than this, and often end their life via mergers, acquisitions, or corporate bankruptcies. Recall that none of such events should be possible according to the GBM model, again suggesting that it contradicts the reality.\footnote{While this only becomes evident in a long run, it does 
\emph{not} mean that the GBM model is only `asymptotically wrong' rather than being `qualitatively wrong'. As I argued above, a fictitious ever-growing force equal to the negative gradient of the GBM potential is absurd on the whole positive semi-axis $ X \geq 0 $, that is, at each time moment. On the other hand, the simplicity of the GBM potential (\ref{inverted_pot_GBM}) does not justify invoking of an asymptotic analysis to identify regions of the state space $ X \geq 0 $ where the model becomes `too' wrong - it is wrong everywhere.}
 Also note that it would not be fair to use a long history of market index portfolios such as e.g. Dow Jones or S\&P500 as an evidence of an average exponential long-term growth for individual stocks. 
Due to the fact that the composition of such market index portfolios continuously changes, it embeds a survivorship bias that allows it to ignore the fact that stocks can default, and proceed away with the implicit assumption that stocks are immortal.    

An analogy with the Inverted World mentioned in the introduction should become more transparent to the reader at this point. An unlimited \emph{fall} in an unbounded potential 
 (\ref{inverted_pot_GBM}) describing a small noise dynamics of the GBM model and all its descendants can only be perceived as an exponential \emph{growth} 
 only if the observer is somehow `inverted' as well. When viewed from the perspective of the Langevin dynamics, an unbounded average exponential \emph{growth} of assets according to the GBM model turns out to be an unbounded \emph{fall} in the inverted parabolic potential. This is obviously a catastrophic scenario for most models in physics except dedicated models designed to describe short-lived unstable systems (e.g. in some  cosmological models).
 As an unbounded exponential expansion never occurs in most of other natural systems known to physics, real `physical' markets should have mechanisms that eventually stop such an unbounded expansion. 
 As I argue next, such stabilization can arise from taking into account interactions and non-linearities in the market dynamics.

\subsection{Stabilization of dynamics by non-linearities}
\label{sect:Linear_vs_nonlinear}

Both the classical GBM model and a majority of models used in quantitative trading\footnote{Excluding more specialized models specifically addressing market impact, for example for optimal stock execution.} are {\it linear} models, in the sense that they have a linear (or constant) drift. A linear or a constant specification of a drift term for stock dynamics might appear a simplest reasonable choice, given that a drift is harder to measure at small time steps 
$ \Delta t $ than a diffusion term\footnote{This is because a drift and diffusion term scale as $ O \left(\Delta t \right) $ and $ O \left( \sqrt{ \Delta t} \right) $, respectively, therefore the second term dominates when $ \Delta t \rightarrow 0 $.}. 

As is known in physics, linear dynamic systems can typically be only considered \emph{approximations} to real-world dynamics of natural systems, which are 
often {\it non-linear}.  Non-linearities capture interactions in physical systems, that could be produced either by interactions between different elements of a system, or interactions with some external potential. In particular,  within the Langevin approach, the most common approach to incorporate complex interactions in a physical system is to consider more complex potentials than a harmonic oscillator potential. One popular choice are potentials expressed as polynomials in a state variable $ X $. While the use of a general polynomial potential can be justified as a Taylor expansion of an arbitrary potential, for most systems encountered in statistical and quantum physics it usually suffices to consider polynomial potentials up to the fourth degree \cite{Landau} (see also references in \cite{QED}).
   
One of the most popular non-linear potentials describing many systems in physics is the so-called quartic potential 
\beq
\label{pot_4}
U(X) = - \frac{1}{2} \theta X^2 + \frac{1}{3} \kappa X^3 + \frac{1}{4} g X^4,
\eeq
where parameters $ \theta $, $\kappa $ and $ g $ may depend on time through their dependence on various predictors, but can be taken constant in a simplest case.
Obviously, if we set $ \kappa = g = 0 $ and $ \theta = \mu $, this potential recovers the GBM potential (\ref{inverted_pot_GBM}). Otherwise, for 
non-zero values of  $ \kappa, \, g $, the GBM potential (\ref{inverted_pot_GBM}) can serve as a local approximation, valid for small values of $ X $, to the non-linear potential (\ref{pot_4}). 

The fact that linear models such as the GBM model are only approximations to more general non-linear models that incorporate market impact, transaction costs etc. is well known in the literature.  However, there exists a common belief among both market practitioners and academics that such non-linear effects are only important for large players who `move the market', while for small trades the standard linear models that neglect market impact can still be used.     

I propose that even for classical quantitative finance models such as the GBM that do \emph{not} assume a large trader, capturing non-linear effects of market interactions is critically important for constructing more reasonable models that would not produce the absurd ever-accelerating unbounded decay in the inverted parabola potential
 (\ref{inverted_pot_GBM}) of the GBM model. 
 
 Indeed, assume for a moment that the quartic potential (\ref{pot_4}) is a `right' model of the world (I will argue later in favor of this choice, so that the example is not hypothetical). When $ X $ is sufficiently small, the cubic and quartic terms can be neglected, and upon setting $ \theta = \mu $, we recover the GBM potential 
 (\ref{inverted_pot_GBM}) as a small-field approximation to the quartic potential  (\ref{pot_4}).  Depending on model parameters, the behavior of the potential can match the quadratic approximation well in a parametrically wide range of the price $ X_t $. Clearly, if we set parameters  $\kappa $ and $ g $ to zero \emph{exactly}, then the two potentials are identical on the whole semi-axis $ X \geq 0 $.
   
%

\begin{figure}[ht]
\begin{center}
\includegraphics[
width=180mm,
height=50mm]{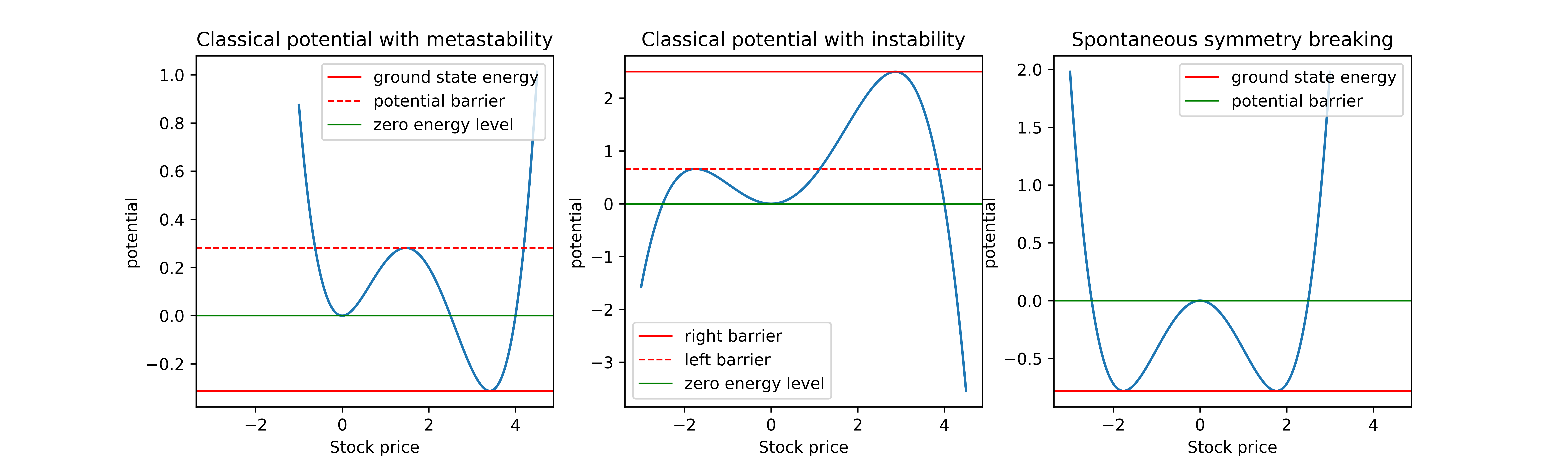}
\caption{Under different parameter choices in the quartic potential $ U(x) $ of Eq.(\ref{pot_4}), it can take different forms. A stable state of the system corresponds to a minimum of the potential. The potential on the left describes a metastable system with a local minimum at zero and a global minimum at $ x = 3.3$. 
If a particle is initially released near the global minimum, most of the time it will experience a small diffusive relaxation towards the global minimum, 
which, with a small probability, can be replaced at each instance by a large sudden jump across the potential barrier separating the two minima. 
The minimum at zero corresponds to the default state. For the potential in the center, the state $ x = 3.3$ becomes unstable, and the state $ x= 0$ is metastable. The potential on the right has two symmetric minima, and the particle can choose any of them to minimize its energy. Such a scenario is called ``spontaneous symmetry breaking'' in physics.}
\label{fig:Classic_potential_shapes}
\end{center}
\end{figure}

On the other hand, for non-vanishing values of parameters $\kappa $ and $ g $ that control, respectively, the cubic and quartic non-linear terms in the potential $ U(x)$, the latter can produce a wide variety of shapes, depending on the values of parameters, as illustrated in Fig.~\ref{fig:Classic_potential_shapes}. 

As was argued in \cite{QED}, it is the potential in the left graph in Fig.~\ref{fig:Classic_potential_shapes} that leads to the most interesting dynamics of a stock market price. Instead of unstable dynamics of the GBM model, with such a potential, dynamics can rather be \emph{metastable}. Such metastable dynamics are different from globally stable dynamics such as e.g. the harmonic oscillator dynamics in that they \emph{eventually} change, though the time for this change to occur may be long, or very long, depending on the parameters. In between of such infrequent transitions, dynamics are approximately equilibrium (stationary) or quasi-equilibrium.
Changes of the dynamics correspond to rare events of transitions between local minima of the potential.  

While an explanation of how this happens will be given momentarily, it is very important to emphasize a critical role of a non-vanishing noise $ \sigma > 0 $ for a realization of a scenario described below. This is because any transitions between different local minima of a potential are only possible when
thermal fluctuations are turned on by allowing for a non-zero  $ \sigma > 0 $. If $ \sigma $ is large, fluctuations become stronger and transitions happen more often, but in the strict opposite limit
$ \sigma = 0 $, any fluctuations die off, and transitions between local minima of the potential are no longer possible. Dynamics obtained in the strict limit 
$ \sigma = 0 $ are qualitatively different from dynamics obtained for non-zero values $ \sigma > 0 $, even though the actual numerical value of $ \sigma $ may be very small numerically. This is the reason why appealing to an exponential bank account law as a justification for a similar average behavior for stocks would be mathematically wrong - as was mentioned in Sect.~\ref{sect_Inverted_parabola_world}, the limit $ \sigma \rightarrow 0 $ is singular (non-analytic).\footnote{A popular example of a non-analytical dependence on a model parameter is given by the function $ f(g) = e^{-A/g} $, where $ g \geq 0 $ is a model parameter, and $ A > 0 $ is a constant. 
Functions of such form are frequently encountered in quantum field theory.
While the value of 
this function is very small for small but non-vanishing values $ g > 0 $, it does not have a Taylor expansion around the point $ g = 0 $, which means that this limit is singular.}     

\begin{figure}[ht]
\begin{center}
\includegraphics[
width=180mm,
height=55mm]{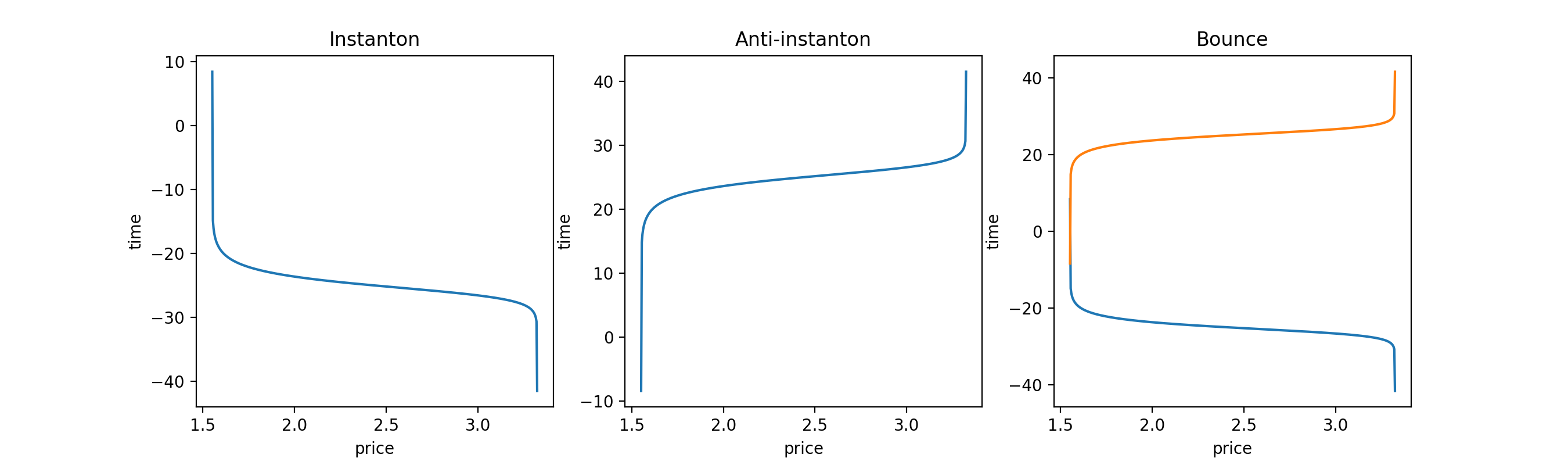}
\caption{Instanton, anti-instanton, and bounce solutions. The instanton hops from the right of a global maximum to the left of it, the anti-instanton proceeds in an opposite order, and the bounce is made of the instanton followed by the anti-instanton.} 
\label{fig:Instanton_and_bounce}
\end{center}
\end{figure}

The potential shown on the left of Fig.~\ref{fig:Classic_potential_shapes}
has a \emph{potential barrier} between a metastable point at the bottom of the local well, and the part of the potential for small values of $ x $, where the motion against the gradient of the potential means a fall to the zero price level $ x = 0 $.
Due to noise-induced fluctuations, a particle representing a stock with value $ x_t $ at time $ t $ placed initially to the right of the barrier, can hop over to the left of the barrier. In physics, solutions of dynamics equations that describe such ``barrier-hopping'' transitions 
are called \emph{instantons}.
The reason for this nomenclature is that the transitions between the meta-stable state and the regime of instability (a ``fall'' to the zero level $ x = 0$) happens almost instantaneously in time. 
What might take a long time though is the time for this hopping to occur: depending on model parameters, the waiting time can in principle even exceed the age of the observed universe. See Fig.~\ref{fig:Instanton_and_bounce}
 for examples of an instanton, anti-instanton (an instanton going backward in time), and a bounce (an instanton-anti-instanton pair, i.e. an instanton followed by an anti-instanton)

In financial terms, an event of hopping over the barrier en route to the zero level at $ x = 0$ corresponds to a corporate bankruptcy (default). As the GBM model
corresponds to the inverted harmonic potential where the point $ x = 0 $ is unattainable, 
corporate defaults cannot be captured by the GBM model. 
In contrast, with the quartic potential  shown on the left of Fig.~\ref{fig:Classic_potential_shapes},  corporate defaults are perfectly possible, and correspond to the instanton-type hopping transitions between different local minima of the meta-stable potential. Note that both the drift (the negative gradient of the potential (\ref{pot_4}))
and volatility vanish at $ X = 0 $. This means the the zero level $ X = 0 $ is an \emph{absorbing} state: once the particle reaches this point, it stays there forever.
This is a highly desirable model behavior as it captures corporate default in a simple diffusion-based stock price model, in a sharp contrast with a failure of the GBM model to produce a defaultable equity model, in addition to unrealistic dynamics for $ X > 0 $.

\section{The ``Quantum Equilibrium-Disequilibrium" (QED) model}
\label{QED}

Unlike the GBM model, the QED model \cite{QED} incorporates capital inflows and outflows in the market, along with capturing their price impact in the model construction.  As I will show below, capturing these phenomena using simple function approximations effectively produces the Langevin dynamics with the quartic potential (\ref{pot_4}), thus offering a plausible mechanism for stabilization of market dynamics by non-linearities as described 
in Sect.~\ref{sect:Linear_vs_nonlinear}. Before providing a mathematical formulation of the model, it is helpful to discuss empirical data.

\subsection{Markets are open systems: importance of money flows and their impact}
 
 Traditional classical finance models such as the GBM model of Samuelson \cite{Samuelson}, the Black-Scholes model \cite{BS}, the CAPM model \cite{CAPM} etc. typically all assume that a market is a closed system that does not exchange cash with outside investors (an ``outside world"). A common assumption for stock dividends often made for modeling stock prices is that any dividends paid by a company are immediately re-invested back into the stock by the shareholders. However, in addition to current investors in a given stock at any point in time, the normal regime of the market is that on average, there is an approximately continuous rate of cash inflows into the market from \emph{new} investors, mainly due to various retirement plans programs. 
 In other words, money is \emph{not} conserved in the market due to continuous inflows (and outflows) of new market participants. 
 
\begin{figure}[ht]
\begin{center}
\includegraphics[
width=90mm,
height=60mm]{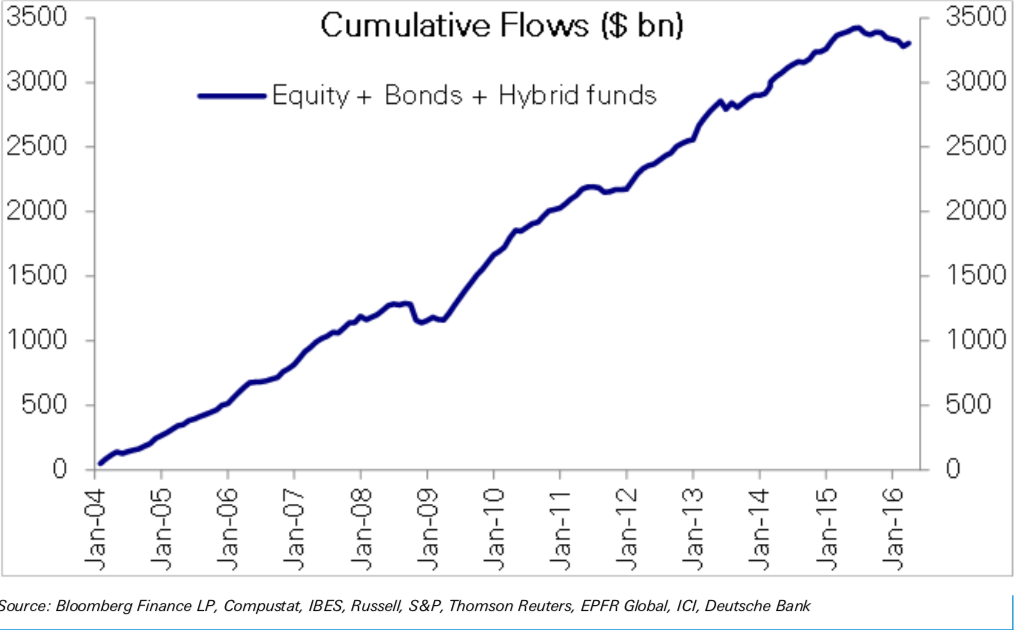}
\caption{Combined inflows into equity, bond, and hybrid funds. The annual rate is approximately constant at the level of \$325bn \cite{DB_2016}.  
} 
\label{fig:fund_flows}
\end{center}
\end{figure}

 Fig.~\ref{fig:fund_flows} demonstrates the dynamics of combined inflows into equity, bond, and hybrid funds \cite{DB_2016}. It shows that on average, there was a steady inflow of around \$325bn annually into the US funds between 2004 and 2016, with a local drop around 2009 as a result of the economic crisis.
 Assuming as a rough estimate that about two thirds of these inflows are invested in stocks, this gives rise to about \$200bn injected every year into the stock market. The main origin of such cash injection are retirement plans of the US workers. 
   
Should an annual injection of \$200bn in the capital market be considered a large or a negligible effect?
The total market capitalization of all stocks in the S\&P500 index is about \$25.5 trillion, or \$25,500bn, so the inflows are of the order of 1\% of the total index value, which may not be a numerically insignificant effect.
In addition, the answer depends on how exactly these inflows are distributed across different stocks. 
If retails or institutional investors are massively driven to invest in a particular ``hot" stock, after a relatively short period of \emph{increased} returns driven ``mechanically" by the momentum, a long term impact of such investor ``crowding" in the stock normally amounts to \emph{diminishing} long-term returns. The latter phenomenon is known as the ``dumb money" effect \cite{Frazzini_2008}. 

Therefore, to model the impact of investors flows and their impact on stock returns, we should simultaneously incorporate two things into the modeling framework, which are both missing in most conventional classical models such as the GBM: capital inflows, and saturation/market friction effects. As we will see next, the QED model incorporates both these effects, and moreover it provides an explanation why these effect are critically important to ensure a long-term stability (or, more accurately, meta-stability, as will be more clear below) of the resulting dynamics, no matter how small these effects may be numerically. 

\subsection{The QED model}
\label{subsect_QED}

 Let $ X_t $ be a total capitalization of a firm at time $ t $, rescaled to a dimensionless
 quantity of the order of one $ X_t \sim 1 $, e.g. by dividing by a mean capitalization over the observation period. 
We consider discrete-time dynamics described, in general form, by the following equations: 
\bea
\label{r_t_one_more}
&& X_{t+ \Delta t} = (1 + r_{t+ \Delta t} \Delta t) (  X_t  - c X_t \Delta t +  u_t  X_t \Delta t )  \nonumber, \\
&& r_{t+ \Delta t}  = r_f + {\bf w}^T {\bf z}_t  + f(u_t)  + \frac{\sigma}{ \sqrt{ \Delta t}} \varepsilon_t, 
\eea
where $ \Delta t $ is a time step, $ r_f $ is a risk-free rate, $ c $ is a dividend rate (assumed constant here),  $ {\bf z}_t $ is a vector of predictors with weights $ {\bf w} $, $u_t \equiv u_t(X_t, {\bf z}_t)$ is a
percentage rate of cash inflow/outflow from outside investors\footnote{Note that here we define cash inflows $ u_t $ as multiples of the total market cap (or equivalently, of the stock market price) $ X_t $, while it was defined in the absolute terms in \cite{QED}.},  $f(u_t) $ is a market impact factor, and $ \varepsilon_t \sim \mathcal{N} (\cdot| 0, 1) $ is white noise.
Here the first equation defines the change of the total market cap\footnote{or, equivalently, the stock price, if the number of outstanding shares is kept constant.} in the time step $ [t, t+\Delta t] $ as a composition of two changes to its time-$t$ value $ X_t $. First, at the beginning of the interval, 
a dividend $ c X_t \Delta t $ is paid to the investors, while  
they also may inject  the amount  $ u_t X_t \Delta t  $ of capital in the stock.
After that, the new capital value $ X_t  + (u_t - c) X_t \Delta t $ grows at rate $ r_{t+\Delta t} $. The latter is given by the second of Eqs.(\ref{r_t_one_more}), where the term $ f(u_t) $ describes the price impact of the money inflow or outflow. In \cite{QED}, we used a simple linear trade impact specification
\beq
\label{impact_linear}
f(u_t) = - \mu u_t
\eeq
where  $ \mu $ is a market impact parameter.  Assuming that $ \mu > 0 $, the chosen sign convention corresponds to a market saturation effect, which may be a proper setting for long-term asset returns.  On the other hand, $ \mu < 0 $ corresponds to a \emph{positive} impact of money inflow $ u_t > 0 $, which may be a relevant setting to describe a short-term impact of 
money inflows due to momentum effects. 
Note that $ u_t $ can be either zero or non-zero, including both positive values and negative values, with a `normal' market corresponding to $ u_t > 0 $.
Another possible specification of the impact function $ f(u_t) $ will be presented below, after we introduce the basic setting.

The reason that the same quantity $ u_t $ appears in both equations in (\ref{r_t_one_more}) is simple.
In the first equation, $ u_t $ enters as a capital injection $ u_t X_t \Delta t $, while in the second equation it enters via the market impact term $ f (u_t)  $ because 
adding capital  $ u_t X_t \Delta t $ means trading a quantity of the stock that is proportional to $ u_t  $. Using a linear impact approximation, this produces the impact
term  $ f(u_t) = - \mu u_t $.

In general, the rate of capital injection $ u_t  $ injected by investors in the market at time $ t $ should depend on the current market capitalization 
$ X_t $ (or current returns), plus possibly other factors (e.g. alpha signals). In \cite{QED}, we considered a simple quadratic choice for $ u_t  $
\beq
\label{determ_u}
u_t  =  \bar{u} +  \phi X_t  + \lambda X_t^2   
\eeq
with three parameters $ \bar{u} $, $ \phi $ and $ \lambda $.\footnote{A slight difference between Eq.(\ref{determ_u}) and a similar formula presented in \cite{QED} is because here we define $ u_t $ as a rate, rather than in absolute terms as was defined in  \cite{QED}. This difference is inessential as it only produces re-defined parameter values of the final model, given by Eq.(\ref{QED}) below, in terms of original model parameters entering equations (\ref{r_t_one_more}) and (\ref{determ_u}).}   
Note that Eq.(\ref{determ_u}) implies that the total money flow $ u_t X_t \rightarrow 0 $ in Eq.(\ref{r_t_one_more}) when $ X_t  \rightarrow 0 $.
This ensures that no investor would invest in a stock with a strictly zero price. Also note that the Eq.(\ref{determ_u})
can always be viewed as a leading-order Taylor expansion
of a more general nonlinear ``capital supply'' function $ u( X_t, {\bf z}_t ) $
that can depend on both $ X_t $ and signals $ {\bf z}_t $. (alternatively, the capital supply $ u $ can be made a function of returns rather than prices \cite{IQED}). Respectively, parameters $ \bar{u} $, $  \phi $ and $ \lambda $ could be slowly varying functions of signals 
$ {\bf z}_t $. Here we consider a limiting case when they are treated as fixed parameters, which may be a reasonable assumption for time periods when an economic regime does not change too much.

Substituting Eq.(\ref{determ_u}) into Eqs.(\ref{r_t_one_more}), neglecting terms $ O(\Delta t)^2 $ and taking the continuous time limit $ \Delta t \rightarrow dt $
we obtain the ``Quantum Equilibrium-Disequilibrium" (QED) model: 
 \beq
 \label{QED}
 d X_t =\kappa  X_t \left( \frac{\theta +  {\bf w}^T {\bf z}_t }{\kappa}  -  X_t - \frac{g}{\kappa} X_t^2  \right) dt +  \sigma   X_t  d W_t,
\eeq
where $ W_t $ is the standard Brownian motion, and parameters are defined as follows: 
\beq
\label{params}
 \theta  =   r_f - c + \bar{u} ,  \; \; \kappa   =   (\mu -1) \phi  , \; \;  g  = (\mu -1) \lambda. 
\eeq   
The dynamics of the QED model is therefore given by the Langevin equation (\ref{Langevin}) with the following potential
\beq
\label{pot_QED}
U(X) =  - \left( \theta +  {\bf w}^T {\bf z}_t \right) X^2  + \frac{\kappa}{3} X^3 + \frac{g}{4} X^4
\eeq   
When the signals $ {\bf z}_t $ are turned off, this is exactly the quartic potential of Eq.(\ref{pot_4}). As discussed in Sect.~\ref{sect:Linear_vs_nonlinear}, 
for some choices of model parameters, this potential leads to stabilization of dynamics around a metastable potential minimum that prevents the stock price from an indefinite growth, while also allowing for corporate defaults (Fig.~\ref{fig:Classic_potential_shapes}). The latter proceed via instanton transitions that correspond to sudden thermally induced jumps over the top of a potential barrier separating different local minima of the potential (\ref{pot_QED}). Instanton solutions in the QED model are illustrated in Fig.~\ref{fig:Instanton_and_bounce}.

It should be noted that the linear impact function (\ref{impact_linear}) may be overly simplistic. Indeed, assuming that $ \mu > 0 $ and $ u_t > 0 $, it implies  that when new money are invested in the stock, it produces an immediate negative impact on the next-period returns. This goes contrary to the presence of momentum effects in the markets that predict that, unless the stock is ``saturated" or ``crowded", an injection of the new money increases the demand and should \emph{increase} rather than decrease returns. As shown in \cite{IQED}, instead of a linear impact model, a quadratic model
with time-dependent parameters can better capture the 'dumb money' effect \cite{Frazzini_2008} that predicts that an initial flow into a stock should increase expected returns, but a continuous buildup of inflows into the stock leads (crowding) leads to diminishing long-term returns. 
With such a choice of the impact function,  we can retain only the linear term in Eq.(\ref{determ_u}) to come up with the same QED dynamics (\ref{QED}) and the
potential (\ref{pot_QED}), albeit with different expression for parameters $ \theta, \kappa $ and $ g $ in terms of original model parameters entering 
Eq.(\ref{r_t_one_more}).

\subsection{QED model and instantons: non-perturbative finance}
 
For some physical systems, non-linearities  can be handled approximately, by treating them as small perturbations around a linear regime, using e.g. a perturbation theory in a small parameter that quantifies the stength of non-linearity.
However, in many other cases arising in the natural sciences, non-linearities should be treated as key ingredients of the dynamics.

For example, non-linearity is critical for self-organizing systems which cannot be described using a perturbation theory around a linear regime.
Another well-known example is provided by instantons -  barrier transition phenomena in statistical and quantum physics  discussed above. Probabilities of such barrier transitions cannot be obtained at any finite order of a perturbation theory in a small parameter controlling the non-linearity. They are examples of so-called {\it non-perturbative} phenomena. While 
instantons and other non-perturbative phenomena are very important in many models of statistical physics and quantum field theory\footnote{Including e.g. quantum chromodynamics (QCD), the modern theory of strong interactions. To explain the very existence of protons and neutrons, QCD needs to go beyond perturbation theory.}, they are not traceable using tools of perturbation theory, see e.g. references cited in \cite{QED}.

Similarly, instantons in the QED model (see Fig.~\ref{fig:Instanton_and_bounce}) are non-perturbative phenomena in parameters $ \kappa, g $, and thus could not be seen at any finite order of a perturbation theory constructed around a strict limit $ \kappa = 0, \, g = 0$ of the QED dynamics. As in this strict limit the QED model would be identical to the GBM model, this means that while the latter could formally be considered as a 'baseline', unperturbed model for construction of such a perturbative expansion, instantons (and hence corporate defaults) would be entirely lost in such a scheme. Non-perturbative methods to compute instanton-induced transition probabilities associated with probabilities of corporate defaults are presented in \cite{QED}. As was illustrated in \cite{QED}, this enables a simultaneous calibration of the QED model to equity and credit markets, by a joint fit to equity returns and credit default swaps (CDS) spreads. In its turn, it enables using data from credit markets to produce information on a long-term equity returns. The QED model is therefore a first defaultable equity model that captures corporate defaults  without introducing additional degrees of freedom such as hazard rates.  
  
\section{Summary}
\label{sect:Summary}

To summarize, starting with Samuelson's GBM model, many models used by practitioners for modeling stock prices and derivatives prices, such as local or stochastic volatility models, relied on the assumption of a linear (and typically positive) drift of a price process, or equivalently a constant drift of a log-price process. In this paper   
I showed that, when interpreted in physics terms, these models describe an oscillator with a \emph{negative} mass (or equivalently a \emph{particle in an inverted parabolic potential}) subject to noise, where differences between specific models amount to different ways of modeling noise.
This makes them all models of stochastic dynamics in an \emph{unstable} potential, and conflicts with conventional ways of analysis of natural systems in physics where models typically describe fluctuations around some \emph{stable or metastable} state. A \emph{qualitatively wrong behavior} describing an unlimited fall in such an unbounded potential is obtained as a result.  Samuelson's solution of the problem of negative prices in the ABM model of Bachelier is unsatisfactory as it leads to a conflict with basic physics. 

I argued that such a pathological behavior can be avoided if the market is modeled as an open system with a possible exchange of money with an outside world, along with a price impact of the new money on stock prices.  For a single-stock market,  this produces a simple non-linear two-parametric extension of the GBM model, 
with new parameters $ \kappa, g $,  called "Quantum Equilibrium-Disequilibrium" (QED) model \cite{QED}. The QED model formally transforms into the GBM model in the limit  $ \kappa, g \rightarrow 0 $. With non-zero parameters, it produces a qualitatively different behavior: while the GBM model describes \emph{unstable} dynamics, the QED model describes \emph{metastable} dynamics where a diffusive relaxation to a metastable state is followed by a rare large negative move describing a transition to a distressed state or corporate bankruptcy. Such rare large moves are due to noise-induced solutions of the model called instantons. Similarly to instantons in physics, instantons in the QED model are \emph{non-perturbative} phenomena: they cannot be seen in a perturbative expansion of the model that could be attempted when parameters  $ \kappa, g $ are small but non-zero. In particular, instanton disappear in the strict 'GBM limit' $ \kappa = g = 0 $. 

The QED model offers a few important theoretical insights. While classical financial models have traditionally focused on modeling volatility while keeping simple linear assumptions of the drift, the QED model suggests that the drift should instead be \emph{non-linear}, and should be identified \emph{prior} to analyzing 
volatility patterns. 
In particular, it would be interesting to reconsider various stochastic volatility and `rough volatility' models after fixing the drift function alone the lines suggested in this paper. 

The QED model can be extended along multiple dimensions. In particular, it can be extended to a market with multiple assets, producing the IQED (``Interacting-assets QED") model \cite{IQED}. Other possible extensions can make the quartic potential random - for example, by allowing a dependence of parameter $ \theta $ on signals $ {\bf z}_t $ as in Eq.(\ref{pot_QED}). This may make the dynamics of the model more realistic and avoid a possible negative long-term drift that might be obtained in the model if the potential is kept static. Clearly, for applying the model for a long term modeling, one should better make some or all model parameters dependent on 
signals ${\bf z}_t $, assuming that the latter carry information on a contemporaneous market environment. Proceeding in such way would effectively promote the potential to a random quartic potential.   

Obviously, an important practical question is that assuming that the QED model is a `right' model, how important are non-perturbative effects implied by the model? Can we still rely on traditional financial engineering models that are all based on the assumption of a linear or constant drift? If yes, when can we still rely on them? Such questions can (and should) be answered for any particular stock market and any traditional model by comparing results obtained with that model
versus the QED model. In general, non-linear and non-perturbative effects are \emph{not} expected to be critically important for small local price fluctuations. However, it is a comparison with a more general non-linear model such as the QED model that 
should answer the question about a range of prices and times where traditional linear models can still be used.  
The QED model could also be used for option pricing, and its predictions could be analyzed and compared with traditional models both numerically and analytically using various approximations. Results of such analysis will be presented elsewhere.   


%
%

\end{document}